# Detecting multiple chirality centers in chiral molecules with high harmonic generation

**Ofer Neufeld[1,2*], Omri Wengrowicz[2], Or Peleg[2], Angel Rubio[1,3], and Oren Cohen[2*]**

[1]*Max Planck Institute for the Structure and Dynamics of Matter and Center for Free-Electron Laser Science, Hamburg, 22761, Germany.*
[2]*Technion – Israel Institute of Technology, Physics Department and Solid State Institute, Haifa, 3200003, Israel.*
[3]*Center for Computational Quantum Physics (CCQ), The Flatiron Institute, New York, NY, 10010, USA.*
**Corresponding authors E-mails: ofer.neufeld@gmail.com, oren@technion.ac.il*

**Characterizing chirality is highly important for applications in the pharmaceutical industry, as well as in the study of dynamical chemical and biological systems. However, this task has remained challenging, especially due to the ongoing increasing complexity and size of the molecular structure of drugs and active compounds. In particular, large molecules with many active chirality centers are today ubiquitous, but remain difficult to structurally analyze due to their high number of stereoisomers. Here we theoretically explore the sensitivity of high harmonic generation (HHG) to the chirality of molecules with a varying number of active chiral centers. We find that HHG driven by bi-chromatic non-collinear lasers is a sensitive probe for the stereo-configuration of a chiral molecule. We first show through calculations (from benchmark chiral molecules with up to three chirality centers) that the HHG spectrum is imprinted with information about the handedness of each chirality center in the driven molecule. Next, we show that using both classical- and deep-learning-based reconstruction algorithms, the composition of an unknown mixture of stereoisomers can be reconstructed with high fidelity by a single-shot HHG measurement. Our work illustrates how the combination of non-linear optics and machine learning might open routes for ultra-sensitive sensing in chiral systems.**

## 1. INTRODUCTION

Chirality is a ubiquitous property in nature. It manifests in a plethora of physical and chemical systems ranging from fundamental particles [1], to galaxies [2], and materials [3–5]. In biology and medicine, chirality plays an essential role in determining the body's physiological response to drug administration. This is a direct consequence of the inherent chirality of biologically active molecules such as DNA and amino acids [6,7]. A major task in modern drug development is thus to identify, classify, and separate, the different chiral constituents of molecules to a high precision [8,9]. In fact, a new drug cannot receive FDA approval until a sufficiently pure mixture has been synthesized and characterized [10]. However, this task has become exponentially difficult ever since drug development has moved towards larger biological molecules and chemical species that possess a high number of active chirality centers (stereocenters, see illustration in Fig. 1). 'Multi-center' molecules have a total of $2^N$ stereoisomers (isomers of the same chiral molecule that differ only by the handedness around each chirality center) to be separated, where $N$ is the number of stereocenters [9]. Notably, characterization of chirality in multi-center molecules is a very challenging task, because no single standard all-optical measurement is sufficient. For instance, circular dichroism (CD) spectroscopy [11] (which is the workhorse of the pharmaceutical industry) cannot characterize a mixture of four or more stereoisomers (i.e N>1), because signals from different stereoisomers can either cancel out or be indistinguishable. Instead, current approaches rely on an ad-hoc combination of multiple measurements from both optical and chemical methods, which are not always transferable and can be very time-consuming.

In this context, recent years have seen an explosion of all-optical chirality characterization methods of enhanced sensitivity, including second-order nonlinear optical methods [11–16], enantio-specific state transfer [17], and methods based on high harmonic generation (HHG) [18–21]. However, in all of these studies, only a single chirality center was considered, and it is not clear how, or if, these methods can be extended to probe the structure of multi-center molecules.

Here we theoretically demonstrate all-optical chiral spectroscopy for multi-center molecules utilizing strong lasers and HHG. We employ bi-chromatic laser fields that are non-collinearly aligned and possess an optical property called 'local-chirality' [21] (which has been recently shown to be effective for probing single-center chiral molecules [21–23]). We find that this configuration leads to large chiral signals between different stereoisomers of molecules with multiple chirality centers. In particular, the extreme non-linear nature of the HHG process imprints, with a high selectivity, unique fingerprints of the handedness and structure of each chirality center onto the harmonic spectra, allowing the stereoisomers to be distinguished by harmonic power measurements. By simultaneously measuring several harmonic orders, this scheme can be applied to characterize the composition of an unknown mixture of stereoisomers with high precision, where the reconstruction is performed in a single-shot measurement that is fed into an algorithm that we developed (either classical or deep learning based).

## 2. METHODOLOGY

### A. High harmonic generation calculations

We begin by describing the methodology used in numerical calculations. Calculations were performed with two sets of benchmark chiral molecules with two and three chirality centers localized on carbon atoms. The two-center chiral molecular system was chosen to be 1-Bromo-2-chloro-1,2-difluoroethane ($C_2H_2BrClF_2$, see Fig. 1), while the three-center chiral molecular system was chosen as 2-Bromo-4-chloro-3-fluoropentane ($C_5H_9BrClF$, see Supplementary Material (SM) for illustration). These molecules are linear alkanes with substituted hydrogens, where each carbon atom with four non-identical substituents forms an active stereocenter. Each molecule is labeled according to the handedness of its stereocenters, e.g. 'RS' refers to the two-center chiral molecule with handedness 'R' about the first center and 'S' about the second (see illustration in Fig. 1).

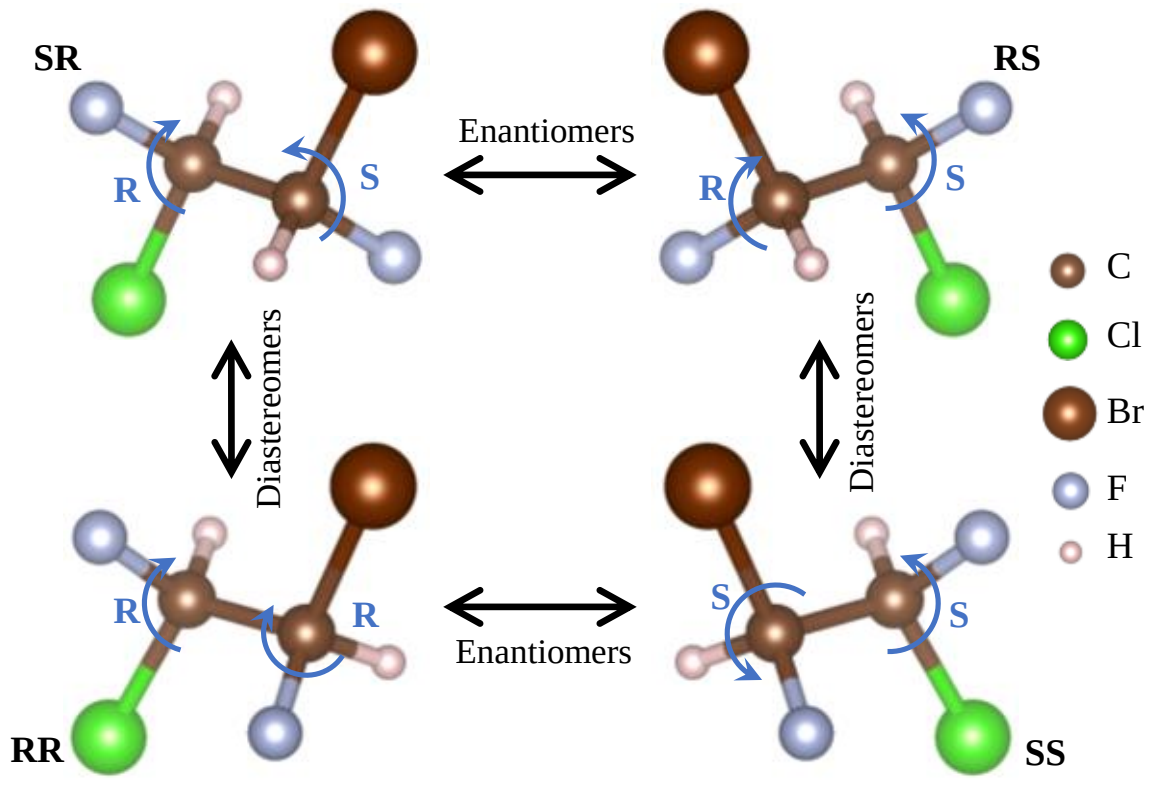

Fig. 1. Schematic illustration of the two-center chiral system showing all four stereoisomers of chiral molecules for $C_2H_2BrClF_2$. The arrows indicate enantiomeric pairs of molecules that are mirror images of one another, as well as diastereomers that are not connected by a mirror symmetry. Labeling of the chiral center is around the two carbons in the chain, respectively, e.g. 'RS' labels a chiral molecule with 'R' configuration around the 1st carbon, and 'S' around the 2nd.

The ground states of the molecules were obtained from density functional theory (DFT) with the real-space grid-based OCTOPUS code [24–26] within the local density approximation (LDA) for the exchange-correlation (XC) functional with an added self-interaction correction (SIC) [27]. Technical details of these calculations are specified in the SM.

For each isomer, HHG calculations were subsequently performed within the single active electron (SAE) approximation for the highest occupied molecular orbital (HOMO) in the length gauge, governed by a time-dependent Schrödinger equation (TDSE) given in atomic units by:

$$i\frac{\partial}{\partial t}|\psi_\Omega(t)\rangle = \begin{bmatrix} -\frac{1}{2}\nabla^2 + V_\Omega(\mathbf{r}) \\ +\mathbf{E}(t)\cdot\mathbf{r} \end{bmatrix} |\psi_\Omega(t)\rangle \qquad (1)$$

where $|\psi_\Omega(t)\rangle$ is the SAE wave function that at $t=0$ was taken as the Kohn-Sham (KS) HOMO orbital, $\mathbf{r}$ is the electron's coordinate, and $V_\Omega(\mathbf{r})$ is the effective potential felt by the SAE that includes both the ionic potential and screening effects from deeper orbitals at a fixed molecular orientation, $\Omega$. $V_\Omega(\mathbf{r})$ was taken as the KS potential of the ground state system:

$$V_\Omega(\mathbf{r}) = -\sum_I \frac{Z_I}{|\mathbf{R}_I - \mathbf{r}|} + \int d^3r' \frac{\rho_\Omega(\mathbf{r})}{|\mathbf{r}-\mathbf{r}'|} + v_{XC}[\rho_\Omega(\mathbf{r})] \qquad (2)$$

where $Z_I$ is the charge of the $I$'th nuclei and $\mathbf{R}_I$ is its coordinate, $v_{XC}$ is the XC potential that is a functional of $\rho_\Omega(\mathbf{r})$, the ground state electron density for that particular orientation. Effectively, the above numerical procedure is equivalent to performing time-dependent DFT calculations, but where the KS potential is frozen to its ground state form, leading to the independent particle approximation, and where only the HOMO KS state is propagated in time. $\mathbf{E}(t)$ in eq. (1) denotes the time-dependent laser electric field of the ω-2ω bi-chromatic non-collinear laser configuration:

$$\mathbf{E}(t) = E_0 A(t) \mathrm{Re}\{e^{i\omega t + i\eta}\hat{\mathbf{e}}_1 + \Delta e^{2i\omega t}\hat{\mathbf{e}}_2\} \qquad (3)$$

where $E_0$ is the electric field amplitude of the ω beam, Δ is the amplitude ratio between the beams, $\eta$ is a relative phase, $\hat{\mathbf{e}}_{1,2}$ are unit vectors along the polarization direction of each beam (each beam is circularly-polarized and counter-rotating, but propagating with a small opening angle α w.r.t the $z$-axis), $A(t)$ is a dimensionless envelope function (taken to be trapezoidal with two-cycle rise and drop sections and a four-cycle flat-top). Note that we have applied the dipole approximation and neglected all spatial degrees of freedom in $\mathbf{E}$(t). Calculations were thus performed microscopically at the focus of both beams where they overlap spatially and temporally. While this approximation is not strictly valid for non-collinear beams (since the phase of $\mathbf{E}$(t) can change across the interaction region), such issues can be avoided either by employing a thin interaction region, or by controlling the relative phases between the beams [21].

Equation (3) describes two noncollinear circularly polarized laser beams of frequencies ω and 2ω, which are focused together into a randomly-oriented chiral media (as illustrated in Fig. 2). For $\alpha=0$ one recovers the collinear planar form of $\mathbf{E}$(t) (i.e. the field is polarized in the $xy$ plane) and $\mathbf{E}$(t) reduces to a counter-rotating bi-circular field [28–30], as has been utilized to generate circularly-polarized high harmonics [31,32]. For $\alpha\neq0$ one obtains a locally-chiral field with a nonzero degree of chirality (DOC) [33]. This field has a unique property of being distinct from its mirror image within the dipole approximation, which leads to strong chiral signals in the nonlinear optical response [21].

The TDSE in eq. (1) was solved within OCTOPUS code (further technical details are found in the SM) for each molecular orientation w.r.t the driving laser field. From the obtained $|\psi_\Omega(t)\rangle$ we calculated the time-dependent dipole acceleration:

$$\mathbf{a}_\Omega(t) = \frac{\partial^2}{\partial t^2}\langle\psi_\Omega(t)|\mathbf{r}|\psi_\Omega(t)\rangle \qquad (4)$$

We performed orientation averaging for each isomer to describe randomly oriented chiral media, where the angle Ω was integrated over with trapezoidal weights on an Euler grid in a method similar to that used in ref. [23], with a total of 208 non-equivalent angular orientations (405 total orientations). From this integration we obtained the dipole acceleration of each randomly oriented molecular isomer:

$$\mathbf{a}(t) = \int d\Omega\, \mathbf{a}_{\mathbf{\Omega}}(t) \qquad (5)$$

, and the HHG spectra was obtained as a Fourier transform of $\mathbf{a}(t)$. For a mixture of different stereoisomers, the HHG spectra was calculated as a coherent sum of the spectra from all isomers in the solution with appropriate weights (according to their individual concentrations).

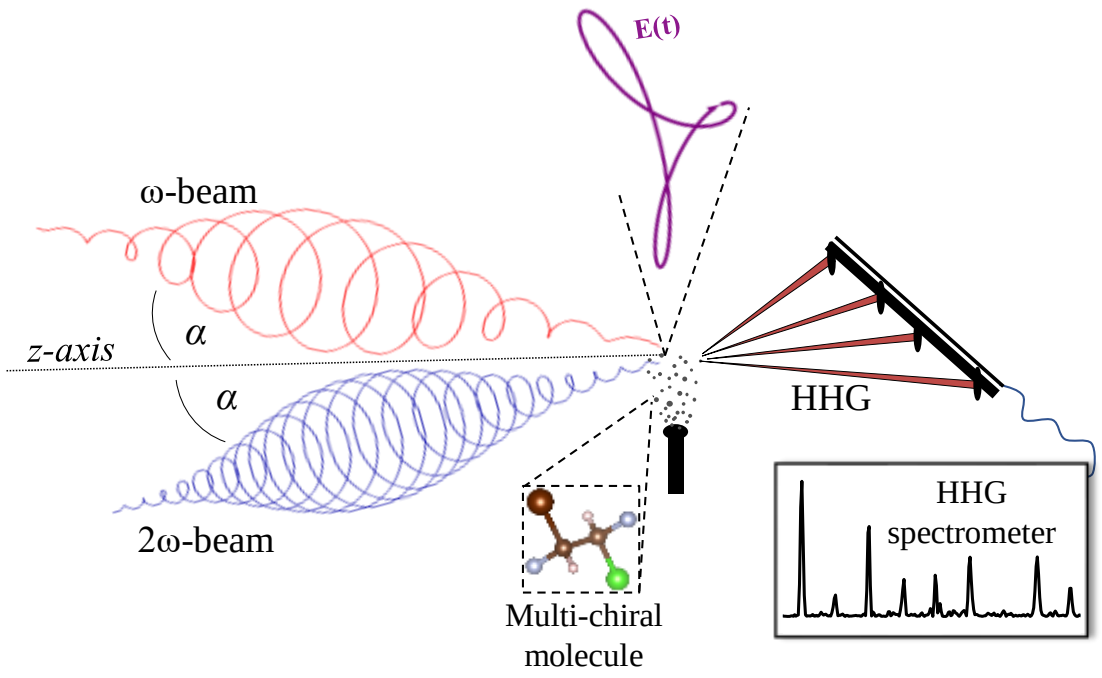

Fig. 2. Schematic illustration of the ω-2ω biochromatic driving field and suggested experimental set-up for chiral mixture reconstruction. The field is comprised of two strong laser beams that are focused together into a mixture of randomly oriented chiral molecules with several stereoisomers (the mixture can either be in liquid or gas phase). The beams have an opening angle (2α) relative to their propagation axes and are circularly polarized in order for the total electric field to exhibit a non-zero DOC [21,33]. High harmonics from the medium are emitted in different angles in space and their powers and polarizations are measured, and then inputted into an algorithm that reconstructs the composition of the mixture.

### B. Chirality reconstruction algorithms

In order to reconstruct the compositions of unknown mixtures of isomers, we developed two algorithmic approaches. The first is based on a classical steepest descent approach, while the second relies on deep learning. Within both approaches, the measured HHG yields for several harmonic orders from a mixture of stereoisomers are inputted into the algorithm. The algorithms output the reconstructed weights for the concentrations of each isomer in the mixture, where it is assumed that the mixture only contains stereoisomers of that particular species (though this constraint could be relaxed in future works). We have simulated many random realizations of mixtures, and quantified the success of the method by comparing the reconstructed weights with the ground truth. The robustness of the method was tested by reconstructing mixture compositions while including random noise in the input data.

We now specify the details of both approaches. The classical approach was implemented in MATLAB [34]. It searches for the optimal molar concentrations of each constituent in the mixture by minimizing a target function defined as the difference between the measured yields, and the reconstructed yields. For each harmonic order we define the target function:

$$f_{tar}{}^{(h)}(\mathbf{c}) = I_{mix}{}^{(h)} - \left|\sum_{j=1}^{M} c_j \sqrt{I_j{}^{(h)}}\, e^{i\phi_{1j}^{(h)}}\right|^2 \qquad (6)$$

where the molar concentrations are labeled as $c_j$ (forming the vector $\mathbf{c}$), "$j$" is the index of the stereoisomer running from 1 to $2^N$ (N is the number of chirality centers in the molecule and we denote $M=2^N$). $I_{mix}{}^{(h)}$ is the input integrated yield of the $h$'th harmonic from the unknown mixture, which is taken from the HHG simulations. $I_j{}^{(h)}$ is the integrated intensity of the $h$'th harmonic from the $j$'th isomer, and $\phi_{1j}^{(h)}$ is the relative phase of the $h$'th harmonic between the 1'st and the $j$'th isomer (where $\phi_{11}^{(h)} = 0$). For the classical steepest descent approach, both of these quantities are assumed to be known reference data that the algorithm has access to. In an experimental set-up, this reference data can be obtained directly from HHG measurements from the pure isomer compounds, as well as 50/50 mixtures of all isomer pairs (while here they are taken directly from the HHG calculations). An overall target function is defined by summing the errors from all harmonics in a subset K:

$$F_{tar}(\mathbf{c}) = \sum_{h\in K} \left|f_{tar}{}^{(h)}(\mathbf{c})\right| \qquad (7)$$

The molar concentrations formally uphold the following constraints that are applied in the algorithm: $1 = \sum_{i=1}^{M} c_i$, which is used to determine: $c_M = 1 - \sum_{i=1}^{M-1} c_i$ (reducing the number of parameters to reconstruct from M to M-1). Also, each $c_i$ upholds: $0 \le c_i \le 1$. This approach essentially searches for a global minimum for $F_{tar}(\mathbf{c})$ with respect to M-1 parameters; thus, reasonable solutions are expected when utilizing at least M-1 harmonic orders for the reconstruction process.

The deep learning approach is implemented in Pytorch [35] and is comprised of two modules. The first module is denoted as HHGNet, which models the measurement process. It is a non-parametric module (non-trained) that is used to generate the expected yields from a given simulated mixture of isomers. The simulated spectrum is converted to logarithmic scale and normalized in the range [-1,1], and is then fed as input into the second module, denoted as MuChiNet. This module is trained to estimate the molar concentrations corresponding to the input yields. It is comprised of two 1D convolution layers, each followed by a MaxPool and a ReLU nonlinearity, and four fully connected layers, each followed by a ReLU nonlinearity [36]. The output is normalized by a softmax layer such that the sum of concentrations gives the physical 100%, which is comparable to the ground truth. The Adam optimizer [37] is used to train the network over 1000 random simulated samples (which are fully-independent from the subsequently used testing set). Crucially, within this technique the relative phases for each harmonic order between isomers ($\phi_{1j}^{(h)}$) are not used as input, significantly simplifying the experiment.

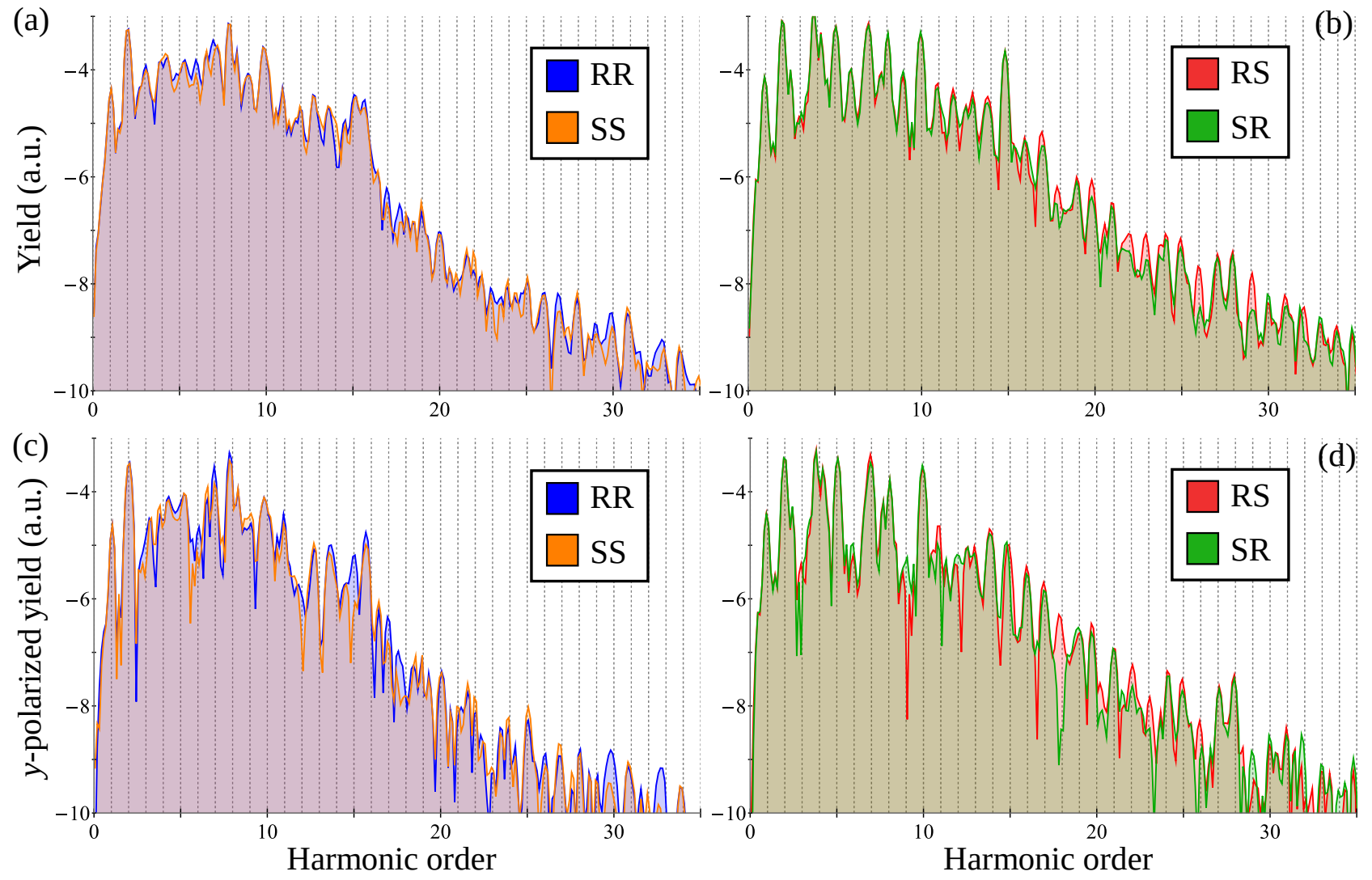

Fig. 3. Extreme nonlinear response and harmonic emission from two-center chiral system, $C_2H_2BrClF_2$. (a,b) HHG power spectra from pairs of enantiomers, showing that the harmonic power is selective to the stereo-configuration of the system. Dashed lines represent integer harmonic orders. (c,d) Same as (a,b) but for *y*-polarized spectral power which shows stronger stereo-selectivity. Plot calculated for the laser parameters: $\lambda$=1400nm, $I_0=10^{13}$ W/cm$^2$, $\alpha=7.5^0$, $\eta=0$, $\Delta=1/\sqrt{2}$.

## 3. RESULTS AND DISCUSSION

### A. Stereocenter sensitivity in HHG

With the approach described above, we investigate HHG from multi-chiral molecules, starting with the model two-center chiral system. Figures 3(a,b) present exemplary HHG spectra from the various stereoisomers of $C_2H_2BrClF_2$ driven by locally-chiral light (with $\alpha = 7.5^0$), where enantiomer pairs are plotted together. The HHG spectra show large enantio-discrimination between pairs of enantiomers, especially for higher harmonics. The size of the discrimination can be quantified by defining a chiral dichroism (CD, not to be confused with circular dichroism) per harmonic order *h* as:

$$CD^{(h)} = 2\frac{I_j^{(h)} - I_k^{(h)}}{I_j^{(h)} + I_k^{(h)}} \qquad (8)$$

where *j,k* indicate indices for the various stereoisomers (e.g. in this case *j=RR,SS,RS,SR*) and $I_j^{(h)}$ is the integrated harmonic yield for the *h*'th harmonic order from the *j*'th isomer. The CD in eq. (8) is normalized from -200 to 200% and is a measure for the sensitivity of HHG to the stereo-configuration. For the particular parameters in Fig. 3 we obtain CDs of ~20% for enantiomeric pairs.

Figures 3(c,d) present similar HHG spectra, but showing only the *y*-polarized emitted power. This simulates a set-up where harmonics are passed through a polarizer before entering the spectrometer, which can enhance the CD substantially – for the particular example in Fig. 3 the CD increases on average by a factor of about 2 after the polarizer. Thus, from this point on we only refer to CD that is obtained with a *y*-polarizer, and only the *y*-polarized HHG spectra is used for reconstructions.

Notably, discrimination between stereoisomers that are not enantiomers but are diastereomers (i.e. are not exact mirror images of one another, see illustration in Fig. 1) is even more efficient. The different structures of diastereomers imprint robust unique features onto the HHG spectra, including the plateau structure and cutoff position. CDs for diastereomers in Fig. 3 for instance are on order of ~100%.

We further analyze the CD between different stereoisomers with respect to the laser conditions. A main degree of freedom here is the opening angle between the beams, α. Fig. 4 shows the CD values *vs.* α for the different harmonic orders between the different isomers. For enantiomer pairs, the CD increases almost linearly with α starting from α=0 (see Fig. 4(a,b)). This behavior is in correspondence with the properties of the driving bi-chromatic laser that vary with α: (i) for α=0 we have a co-planer bi-circular driving field where every third harmonic is suppressed due to symmetry [38,39]. In this geometry the field is not locally-chiral (it has a zero DOC) due to its planar nature, which leads to zero CD. (ii) As α increases the DOC in the driving field increases linearly [33]. Thus, Fig. 4(a,b) shows numerical evidence for the correspondence of HHG chiral signals with light's DOC, which has also been recently reported for photoelectrons [23]. One point to note is that the CD between diastereomers is mostly independent of the beams opening angle (see Fig. 4(c,d)). This indicates that the ability to separate diastereomers in HHG is largely independent of light's chirality, and is instead a consequence of the extremely nonlinear nature of the interaction that is sensitive to the chemical properties of the system (e.g. orbital and potential energy structure). We also note that few harmonics in Fig. 4(c,d) do show strong dependence on α, which is a consequence of interference effects (the electron trajectories are altered with $\alpha$, changing their relative phases).

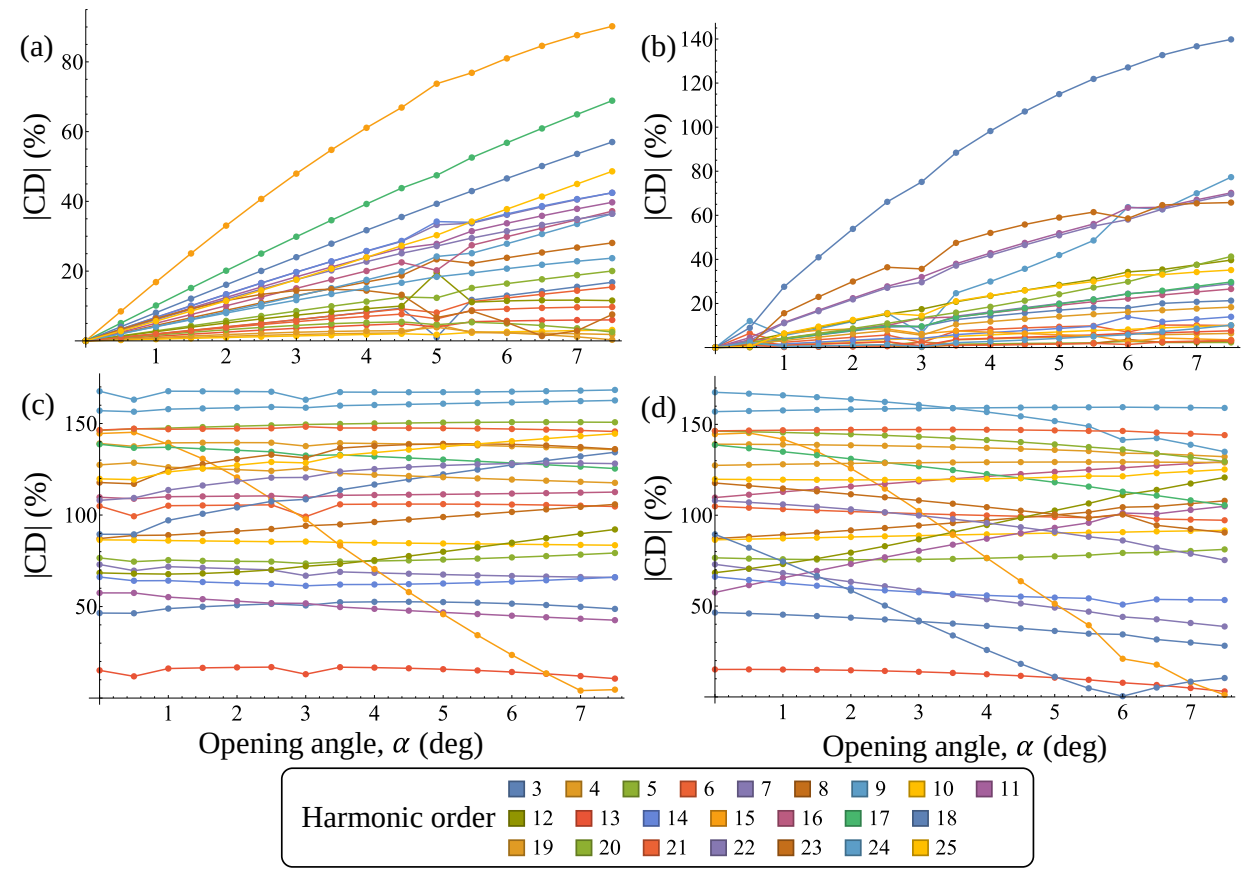

Fig. 4. Chiral dichroism from two-center multi-chiral system *vs.* the bi-chromatic field opening angle, $\alpha$. (a) CD per harmonic order between the enantiomers RR and SS of $C_2H_2BrClF_2$. (b) Same as (a) but for the enantiomers RS and SR. (c) same as (a,b) but for the diastereomers RR and RS. (d) Same as (c) but for the diastereomers RR and SR. Plot calculated for the same laser parameters as Fig. 3, but where $\alpha$ is varied up to $7.5^0$.

Next, we perform calculations in the benchmark three-center chiral molecular system, $C_5H_9BrClF$. Fig. 5 shows exemplary HHG spectra from the various stereoisomers of $C_5H_9BrClF$, which clearly indicates the strong selectivity persists even in molecules with a high number of chirality centers. Fig. 5(e,f) shows that CD values range around 30% for enantiomeric pairs, and 80% for diastereomers, similarly to the two-center system. The fact that each stereoisomer leads to unique fingerprints in the harmonic spectra, as well as very large CDs, opens the possibility for characterization schemes based on HHG, as we will show in the next section.

### B. Reconstruction of chiral mixtures

We have simulated the response of random sets of mixtures of stereoisomers, and employed our developed reconstruction algorithms (classical- and deep-learning-based) to retrieve the molar concentrations of the mixtures. Results for both the two- and three-center multi-chiral molecules are presented in Fig. 6 from both algorithms, where the reconstruction utilizes the *y*-polarized harmonic power that was shown to produce better selectivity (utilizing 20 harmonics in each case). The reconstruction error in a given simulated mixture is calculated as the mean error between the ground truth and the reconstructed concentrations of all stereoisomers. Statistics of errors is obtained for 500 simulated mixtures, from which the mean error and the standard deviation of the error are calculated. Noise was introduced as random errors in the harmonic powers of the calculated HHG spectra from the simulated mixture (i.e. that inputted into the reconstruction algorithm).

Figure 6(a,b) shows mixture concentration reconstruction data from the classical algorithm, which is robust to random noise in the harmonic power data – reconstructions as good as 1% are obtained for noise levels up to 1% for both the two- and three-center multi-chiral molecules. Importantly, the reconstruction quality and noise robustness almost does not change between the two-center and three-center system, indicating that the method should be highly applicable to even larger complex molecular structures.

For optimization purposes, one may also consider optimizing the chosen subset of measured harmonic orders that are fed into the algorithm (the subgroup K in eq. (7)). For instance, one might expect that harmonics that show stronger dichroism between pure enantiomer pairs would increase the signal to noise ratio and reduce reconstruction uncertainties. We have tested various such schemes, but generally observed weak dependence of the reconstruction quality with respect to the harmonic input data, as long as roughly the same number of harmonics were utilized for the reconstruction. On the other hand, by measuring more harmonic orders the reconstruction error can indeed be reduced, as expected (not presented).

We highlight that the stereocenter specificity in Fig. 6 is not possible with standard all-optical methods such as circular dichroism absorption spectroscopy, further emphasizing the importance of the suggested method. One particular example that stresses this difference is a mixture of chiral molecules with equal concentrations for each of the enantiomeric pairs of the stereoisomers, for instance, a mixture comprised of equal parts of all constituents (e.g. 25% for each RR, SS, RS, SR in the two-center system). In this case, the total mixture is racemic, making it completely inaccessible with standard chiroptical techniques regardless of the number of measurements performed. In contrast, our approach successfully reconstructs the concentration of each constituent to a high precision, and works just as well as it does in any other mixture.

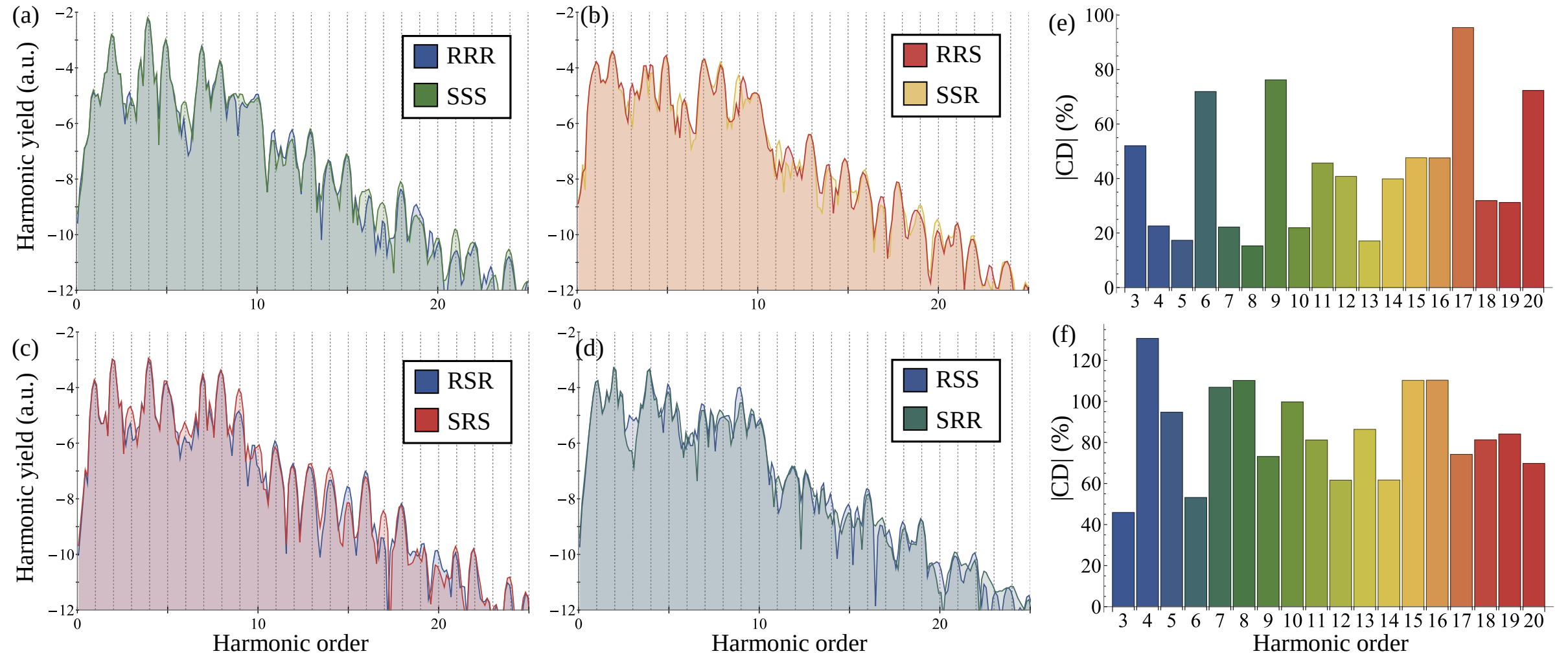

Fig. 5. HHG and chiral dichroism from the three-center multi-chiral system, $C_5H_9BrClF$. (a-d) *y*-polarized HHG power spectra for the various stereoisomers of $C_5H_9BrClF$, where enantiomer pairs are plotted together. (e) |CD| per harmonic order between enantiomer pairs (averaged over all pairs – RRR/SSS, RRS/SSR, RSS/SRR, RSR/SRS). (f) Same as (e) but averaged over all diastereomer pairs. Plots calculated for the same laser parameters as Fig. 3, but for $\lambda$=800nm.

Figures 6(c,d) present similar reconstruction data using the deep-learning algorithm. Reconstructions are generally at the same level of quality, or better, than those obtained with the classical approach for low noise levels. At first glance this may seem like a disappointing result, because one expects the deep learning approach should greatly out-perform the classical algorithms for global minimization. However, we recall that the classical approach in Fig. 6(a,b) assumed that the harmonic phases are known reference data that the algorithm has access to, while the deep-learning approach does not utilize this data. In that respect, the deep-learning scheme has the potential for enhanced realizations in an experimental set-up, provided that one can obtain good quality training data. We also note that the deep-learning approach is more robust over much more noisy data sets, with errors below 4% even up to 50% noise. On the other hand, it tends to lead to an average of 1% errors even in the absence of noise, which is attributed to the lack of phase information (whereas in the classical approach at zero noise the reconstruction equations are exactly solvable, leading to vanishing errors in reconstruction). We emphasize that these results can likely be optimized by improving the algorithmic schemes.

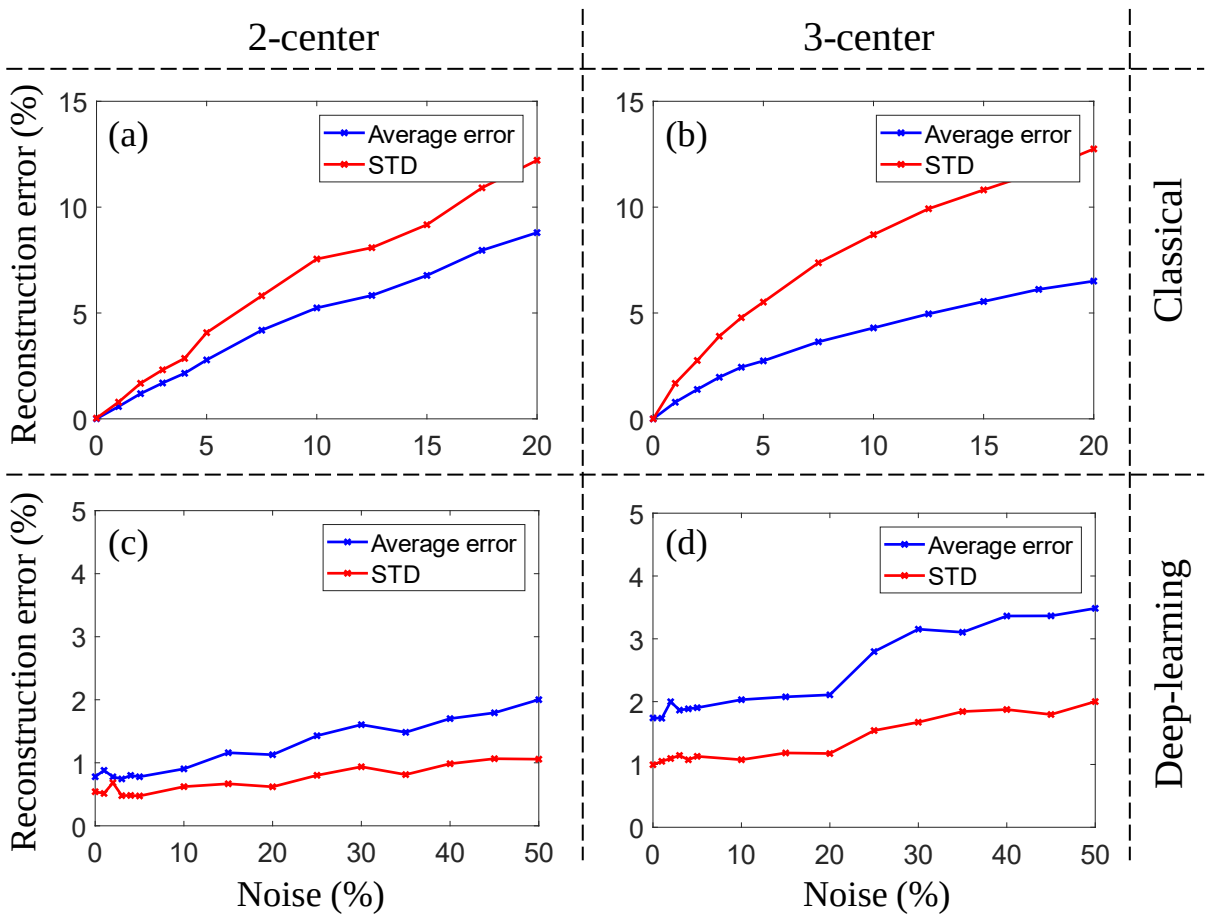

Fig. 6. Algorithmic reconstruction of stereoisomer concentrations in unknown mixtures of chiral molecules. (a) Reconstruction statistics *vs.* noise in harmonic power for two-center multi-chiral molecule using the classical algorithm. Statistics is obtained by simulating 500 random mixtures for every noise level. The absolute error is calculated by averaging the error over all realizations and over all isomers for a given mixture. (b) Same as (a) but for the three-center chiral system. (c,d) Same as (a,b) but with the deep-learning algorithm. Note that the axes scales are different in the classical and deep-learning approaches.

## 4. CONCLUSIONS

To summarize, we have theoretically shown that HHG with 'locally-chiral' light (with bi-chromatic non-collinear lasers [21,33]) is a sensitive probe for molecular chirality and handedness in chiral molecules with multiple stereocenters. The extreme nonlinear nature of the process allows distinguishing between enantiomers and diastereomers, where unique fingerprints of each isomer are imprinted onto the harmonic power and its polarization state. This result allows utilizing HHG for obtaining all-optical characterization of mixtures of chiral molecules (either in gas or liquid phases) to a high precision, in a single-shot measurement. We have simulated the reconstruction process with both classical- and deep-learning-based algorithms, showing that high quality and robust reconstructions can be obtained.

Our results numerically demonstrate the first all-optical method for analyzing chiral molecules with multiple active chirality centers. Looking forward, our work could be used for deriving novel chirality spectroscopy techniques in topological [40,41] and dynamical [42,43] systems based on the combination of extreme nonlinear optics and artificial intelligence.

**Supplemental material**. Supplemental material is available for this paper that includes details on the methodology used in calculations, as well as some additional results.

**Funding.** We acknowledge financial support from the European Research Council (ERC) grant ERC-2015-AdG-694097, and the ERC under the European Union's Horizon 2020 research and innovation programme (819440-TIMP). We also acknowledge financial support from KAMIN program by the Israel Innovation Authority. This work was supported by the Cluster of Excellence Advanced Imaging of Matter (AIM), Grupos Consolidados (IT1249-19) and SFB925. The Flatiron Institute is a division of the Simons Foundation. O.N. gratefully acknowledges the support of the Adams Fellowship Program of the Israel Academy of Sciences and Humanities, support from the Alexander von Humboldt foundation, and support from the Schmidt Science Fellowship.

**Disclosures.** The authors declare no conflicts of interest.

**Data availability.** Data underlying the results presented in this paper are not publicly available at this time but may be obtained from the authors upon reasonable request.

# Supplemental Material: Detecting multiple chirality centers in chiral molecules with high harmonic generation

**Ofer Neufeld[1,2*], Omri Wengrowicz[2], Or Peleg[2], Angel Rubio[1,3], and Oren Cohen[2*]**

[1]*Max Planck Institute for the Structure and Dynamics of Matter and Center for Free-Electron Laser Science, Hamburg, 22761, Germany.*
[2]*Technion – Israel Institute of Technology, Physics Department and Solid State Institute, Haifa, 3200003, Israel.*
[3]*Center for Computational Quantum Physics (CCQ), The Flatiron Institute, New York, NY, 10010, USA.*

## 1. METHODS

DFT calculations were performed with OCTOPUS code [1–3]. The KS equations were discretized in a spherical box with radius 34 Bohr with a Cartesian grid, where molecular centers of mass were centered at the origin. Calculations were performed within the local density approximation (LDA) with an added self-interaction correction (SIC) [4]. The frozen core approximation was used for inner core orbitals, which were treated with norm-conserving pseudopotentials [5]. The KS equations were solved to self-consistency with a tolerance <$10^{-7}$ Hartree, and the grid spacing was converged to 0.4 Bohr. All molecular structures were relaxed <$10^{-4}$ Hartree/Bohr in forces within the LDA.

For TDSE calculations as described in the main text, we utilized a time step $\Delta t$=0.11 a.u. with an imaginary absorbing potential of width 8 Bohr at the boundary. The grid size, absorbing potential, and time step were tested for convergence.

## 2. Three-Center Multi-Chiral System

We present here sketches of the molecular geometries of the utilized three-center multi-chiral system exported in the main text, $C_5H_9BrClF$. Fig. S1 presents the stereo-chemical relationships between the different isomers.

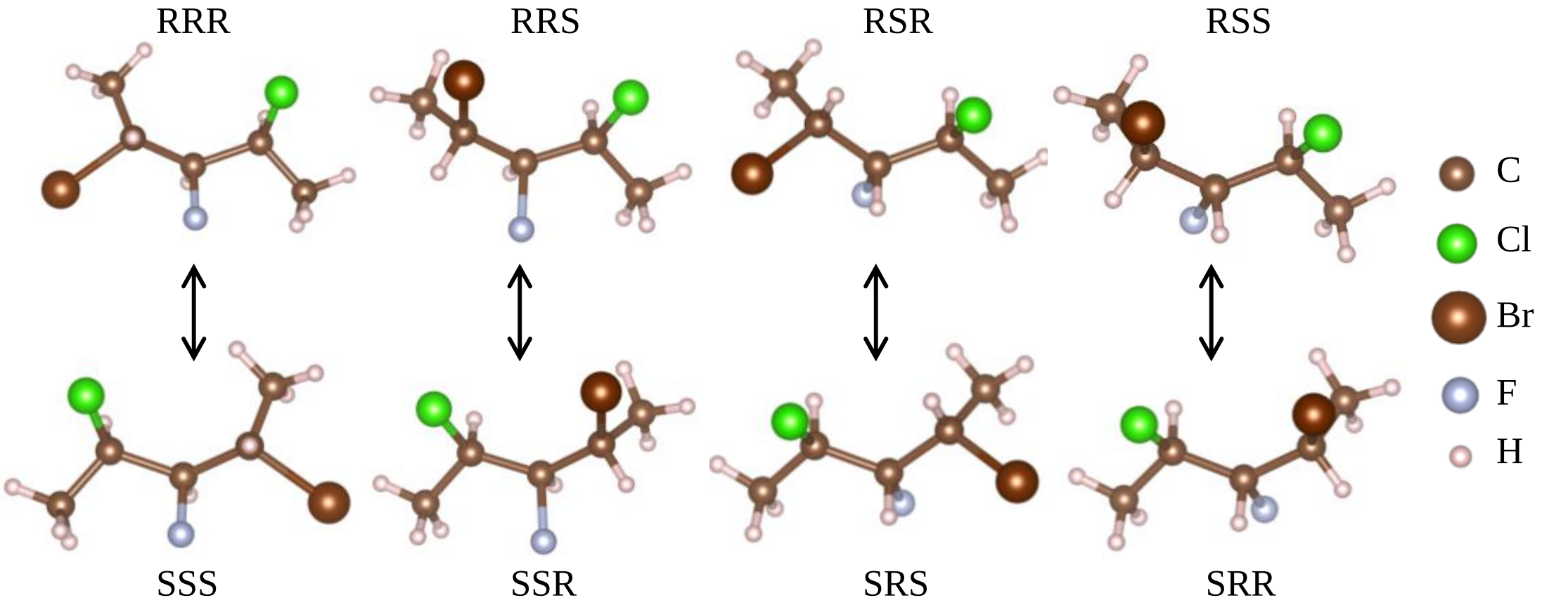

FIG. S1. Schematic illustration of the three-center chiral system showing all eight stereoisomers of chiral molecules for $C_5H_9BrClF$. The arrows indicate enantiomeric pairs of molecules that are mirror images of one another, while diastereomers are not labeled. Labeling of the chiral center is around the three carbons in the chain, respectively, e.g. 'RSR' labels a chiral molecule with 'R' configuration around the 1st carbon, and 'S' around the 2nd, and 'R' around the 3rd.